# Planar Superconductor-Normal-Superconductor Josephson Junctions in MgB$_2$

G. Burnell\*, D.-J. Kang\*, H.N. Lee†, S.H. Moon†, B. Oh†, and M.G. Blamire\*

*\*Dept. Materials Science and IRC in Superconductivity, University of Cambridge Pembroke Street, Cambridge CB2 3QZ, UK*
*†LG Electronics Institute of Technology, Seoul 137-724, Korea*

……………………………………………………………………………………..

**Since the discovery of superconductivity in MgB$_2$[1] considerable progress has been made in determining the physical properties of the material, which are promising for bulk conductors[2-5]. Tunneling studies[6-9] show that the material is reasonably isotropic and has a well-developed s-wave energy gap ($\Delta$), implying that electronic devices based on MgB$_2$ could operate close to 30K. Although a number of groups have reported the formation of thin films by post-reaction of precursors[10-14], heterostructure growth is likely to require considerable technological development, making single-layer device structures of most immediate interest. MgB$_2$ is unlike the cuprate superconductors in that grain boundaries do not form good Josephson junctions, and although a SQUID based on MgB$_2$ nanobridges has been fabricated[15], the nanobridges themselves do not show junction-like properties. Here we report the successful creation of planar MgB$_2$ junctions by localised ion damage in thin films. The critical current ($I_C$) of these devices is strongly modulated by applied microwave radiation and magnetic field. The product of the critical current and normal state resistance ($I_C R_N$) is remarkably high, implying a potential for very high frequency applications.**

Our film deposition technique has been described elsewhere.[12] Briefly, B films were deposited at room temperature on to (0001) sapphire substrates by electron beam evaporation and then ex-situ annealed in a Mg vapour at 850°C for 30 minutes to produce MgB$_2$ films with a final thickness of 100nm and a critical temperature ($T_C$) of 36K. We deposited a 20nm Au film onto the films by dc magnetron sputtering in order to protect the MgB$_2$ from contact with water during subsequent processing. Tracks and contact pads were patterned in the bilayer film using standard photolithography and broad beam Ar-ion milling. The $T_C$ of the patterned tracks was approximately 35K. In order to fabricate the junctions, the film was then transferred to a focused ion beam system (FIB) (Philips-FEI Inc. FIB 200). Chips were wirebonded to enable the resistance of the tracks to be monitored during the FIB milling process.[16] The barrier was defined by writing 50nm wide cuts across the width of tracks using a 4pA 30kV Ga ion beam. The depth of the cut was calibrated by comparing the cut time to that required to completely sever a track; since the Au mills more than order of magnitude faster than the MgB$_2$ it could be ignored in calibrating the cut depth. Compared to other materials used to fabricate SNS junctions using this technique,[17] MgB$_2$ offers the advantage of a relatively low milling rate and hence excellent depth control due to the low atomic mass of its constituents and high melting point.

The devices were measured in dip probes placed in $^4$He storage dewars. The probes are equipped with coils to provide a magnetic field perpendicular to the substrate plane, and with a microwave antenna. The inset to Fig. 1 shows the resistance against temperature for two tracks on the same substrate, one in which no junction was made (a) before placing in the FIB and (b) after the FIB stage, and another track (c) in which a junction had been fabricated. The FIB process clearly has no significant effect on unpatterned films. In case (c) the track with a junction shows an increase in resistance below 125K and a pronounced 'foot' below the main film $T_C$ before becoming completely superconducting at 25K (the absolute resistance is lower than (a/b) due to different track geometries).

Below the $T_C$ of the devices, non-hysteretic, resistively shunted junction (RSJ)-like I-V characteristics were obtained. In Fig. 1 we show the current vs voltage characteristics at several temperatures for a device with a nominal 75% cut in a 2.5µm wide track. Figure 2 shows the temperature dependence of $I_CR_N$ for the same device as Fig. 1. As shown in the inset, $R_N$ (measured from a linear fit to the 20% highest bias in the I-V) is approximately constant at 3.5±0.5Ω over the entire temperature range; this is consistent with the additional resistance upturn in the normal resistance-temperature data extrapolated to 0K and implies a relatively high barrier resistivity of order of 200µΩcm.

Application of an out of plane magnetic field ($B$) to the junctions resulted in a substantial modulation of the critical current. Figure 3 shows $I_C(B)$ at 18K for the same device as in Fig. 2. The apparent incomplete suppression is partly due to the fixed voltage criterion used to determine $I_C$, however there is genuinely a remnant critical current for all applied fields. At lower temperatures the minimum $I_C$ is larger relative to the maximum value which suggests that the barrier is to some degree non-uniform. In a substantially wider junction we observed a rapid SQUID-like $I_C(B)$ modulation attributed to a strongly non-uniform barrier; this is probably a consequence of the surface roughness of the films used.

The $I_C(B)$ modulation clearly differs substantially from the ideal Fraunhofer pattern and is very strongly hysteretic. However, the form is identical to that observed in planar high $T_C$ junctions in which the hysteresis has been successfully modelled on the basis of Abrikosov vortex penetration into the electrodes[18]. Magneto-optic imaging of thin films has shown that polycrystalline $MgB_2$ films such as ours trap flux at very low fields[19]. Mitchell et al.[18] show that in such circumstances the position of the first minimum provides the most reliable estimate of the magnetic penetration depth ($\lambda$). Ignoring flux focusing effects, which are likely be small for a junction of this size, we obtain a value for $2\lambda+d$ of 331nm. Assuming that the electrical length of the barrier is the same as the defined cut width gives a value of $\lambda$ of just over 140nm. This is very similar to estimates obtained elsewhere by different techniques.[20-22]

Complete suppression of the $I_C$ and the formation of Shapiro steps were observed when the device was irradiated with microwaves. We observed the clearest Shapiro steps at a frequency ($f$) of 14.96GHz over a temperature range of 7K to 17K. Above this temperature range the height of the Shapiro steps was too small to observe clearly, whilst below the large $I_CR_N$ product prevented effective coupling of the incident microwave radiation to obtain a complete suppression of the $I_C$. In Fig. 4 we show the amplitude of the $I_C$ and first two Shapiro steps as a function of the incident microwave power for a temperature of 11K. We observed a linear suppression of the $I_C$, as expected for a low reduced frequency ($f/I_cR_n$). At temperatures in the middle of the range we obtained up to four Shapiro steps on the positive and negative branches of the I-V (Fig. 4 inset).

The high value of $I_CR_N$ (1mV at 4.2K) is consistent with that reported for break junctions in bulk $MgB_2$[23]. The temperature dependence of $I_CR_N$ is broadly can be fitted by a solution of the Usadel equations for a structure with a one dimensional non-superconducting weak link in equilibrium with superconducting electrodes for a barrier length ($d$) of the region of 8 times the normal state coherence length ($\xi_{ND}$) implying a value for $\xi_{ND}$ of order of 6nm[24]. For a comparison we also show in Fig. 2 data from a YBa$_2$Cu$_3$O$_{7-d}$ planar SNS junction with the same value of $d/\xi_{ND}$ plotted against reduced temperature[25]: there is excellent agreement between the data consistent with ideal SNS behaviour, but the very different values of $I_CR_N/\Delta$ provides evidence that the two systems are rather different electronically. If we take the modal value of $\Delta$ reported in the literature (5meV)[6-9], $I_CR_N$ for the MgB$_2$ junction is close to the direct prediction of de Gennes,[24] whereas for the YBa$_2$Cu$_3$O$_{7-d}$ junctions it is much lower, implying strong gap suppression in the electrodes.[25] Direct calculation of $\xi_{ND}$ ($(\hbar v_f l/(4\pi kT_c))^{0.5}$) is difficult given the uncertainty of the mean-free path ($l$), but using a value of 4.7x10$^5$ for the Fermi velocity ($v_f$)[2], our value for $\xi_{ND}$ implies that $l$ is of the order of 5nm. A value of 60nm for polycrystalline superconducting MgB$_2$ has been deduced previously, so this value does not seem unreasonable if the ion damage strongly reduces the carrier density in the barrier.

It is clear from the results that our fabrication method which uses a FIB to remove a 50nm wide trench from a track strongly modifies the remaining MgB$_2$ by ion-damage and direct Ga-implantation. To our knowledge this is the first report of Josephson junctions in thin film MgB$_2$. The high $I_CR_N$ product for the junctions fabricated in this way compared to other BCS-superconductor SNS junction techniques, coupled with the much higher $T_C$ make these junctions promising candidates for a range of applications.

## Acknowledgements
This work was supported the UK Engineering and Physical Sciences, and Particle Physics and Astronomy Research Councils and the Korean Ministry of Science and Technology under the National Research Laboratory Project.

## Figure Captions

**Figure 1** Current-Voltage characteristics for 5, 13 and 21K. Inset, the resistance above $T_C$ of a track before placing in the FIB (dashed line), the same track after placing in the FIB (crosses), and the track in which the junction in the main figure was made, after the FIB stage (solid line).

**Figure 2** Critical current-normal state resistance product against reduced temperature for the MgB$_2$ device in Fig. 1 (left scale, solid circles) and a junction fabricated by focussed electron-beam irradiation in YBCO (right scale, open squares)[25]. Inset, normal state resistance of the MgB$_2$ device reported here against reduced temperature.

**Figure 3** Critical current for the same device as in previous figures against applied magnetic field and at 18K: continuous line - increasing field, crosses – decreasing field.

**Figure 4** Critical current (solid line, left scale), and height of the first (solid line) and second (dashed line) Shapiro steps against applied microwave power for the same device as in previous figures and at 11K. Inset, Current voltage characteristics for the device at three representative powers.

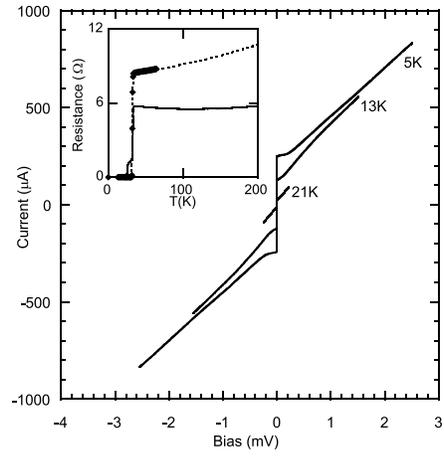

Burnell Figure 1

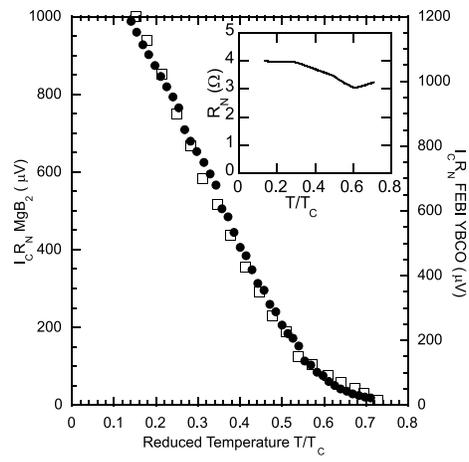

Burnell Figure 2

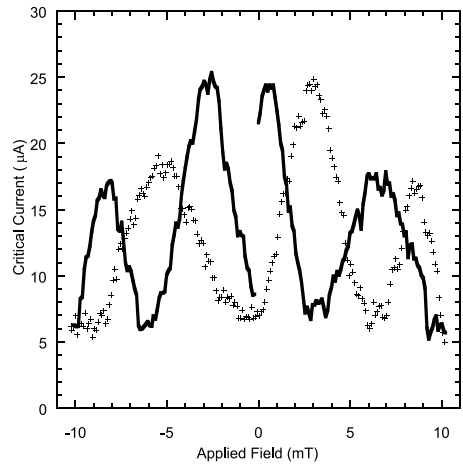

Burnell Figure 3

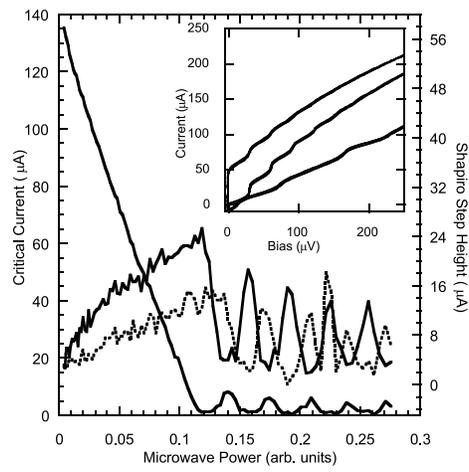

Burnell Figure 4